\documentclass[11pt,a4paper]{article}
\usepackage{sw20aip}

\input tcilatex
\QQQ{Language}{
American English
}

\begin{document}

\author{{\small W.Z. Shangguan \& T.C. Au-Yeung } \\
{\small \textit{School of EEE, Nanyang Technological University, \ Singapore
639798}}}
\title{A Physical Picture of Superconductivity}
\date{}
\maketitle

\begin{abstract}
\baselineskip 24ptA universal mechanism of superconductivity applicable to
``low temperature'' and ``high temperature'' superconductors is proposed in
this paper. With this model of mechanism experimental facts of
superconductors can be qualitatively explained. A function is introduced to
describe the average separation distance between vibrating lattice atoms,
which is crucial for the transition from normal to superconductive state.
However, the most attractive and exciting conclusion that can be derived
from this physical picture, is that given atoms of other element be
successfully sandwiched between ferromagnetic atoms one by one, a
superconductor constructed this way is most likely to have a very high
transition temperature.


\end{abstract}

\baselineskip 24pt

\noindent Since H. Kamerlingh Onnes found superconductivity in 1911\cite
{Onnes}, this unique phenomenon of conductors has perplexed man for nearly a
century, while his understanding about this mysterious property has improved
markedly. In 1957, J. Bardeen, L. N. Cooper and J. R. Schrieffer proposed
BCS theory, which provided a deeper understanding of the microscopic
mechanism of superconductivity\cite{BCS}. However, after Bednorz and
M\"{u}ller suggested that high $T_c$\ superconductors possibly exist in
cuprates in 1986\cite{Bednorz}, the recorder of transition temperature of
superconductors changed constantly and rose drastically. During this period
of time, a wealth of high $T_c$\ superconductivity theories sprung up.\cite
{hi-tc}$^{-}$\cite{Anderson} Yet, the mechanism of superconductivity remains
unsettled, and there is not yet a decisive theory concerning the mechanism
of high $T_c$ superconductivity, though different theories place emphasis on
different aspects and view them from different points. Some authors
postulated that the mechanism responsible for high $T_c$ superconductors may
not be pairing\cite{Anderson}. In addition, it is believed that there should
be no difference between mechanisms of ``low temperature'' and ``high
temperature'' superconductivity. In this paper we propose a universal
mechanism of superconductivity that can be applied to ``low temperature''
and ``high temperature'' superconductors by presenting a physical picture of
conduction electrons and lattice atoms\textbf{. }We view superconductivity
as a consequence of ``vibration harmony'' between conduction electrons and
the vibrating lattice atoms.

It is well-known that the temperature of a material is related largely to
the thermal vibration kinetic energy of its lattice atoms. When its
temperature decreases, the lattice atoms of the material vibrate less
violently. This means that the temperature of the material is the main
factor that affects the vibration amplitude of lattice atoms. As a result,
it is widely accepted that the root cause of superconductivity lies in the
interaction between conduction electrons and lattice atoms of the material,
whose vibration is depicted by energy quanta called phonons. Inasmuch as
superconductor forms when the temperature of the material is sufficiently
low, it is reasonable to infer that the vibration amplitude of the lattice
atoms decreases as the material's temperature decreases, and consequently
the time average separation distance between lattice atoms becomes shorter;
when the vibration amplitude is sufficiently small, electrons can transport
resistlessly among lattice atoms and a superconductor thus forms.

Before we proceed to present the physical picture of superconductivity, it
is instructive to examine the structure of an atom, which is well known as
``solar system''\cite{atom}. That is, the nucleus centers while electrons
move around it, from inner to ouster shell, ending up with Fermi level. The
conduction electrons are those move in the outermost shell.

Let's now take a closer view into a conductor. Fig.1 shows the difference of
lattice atoms array between the normal and superconductive state. When the
temperature of a material is high, its lattice atoms vibrate so vigorously
that some of them are very near while some others quite far, as is
illustrated in Fig.1a. When the temperature of the material is sufficiently
low, however, the lattice atoms vibrate very slightly and array in a better
order, thus the conduction electrons can transfer between atoms without
being scattered and macroscopic resistance exists no longer.\vspace{5cm}

\begin{center}
\textbf{figure 1}
\end{center}

How this transition is realized? For simplicity, we plot only one atom
chain, as shown in Fig.2. Suppose an electron moving clockwise around atom
A, when arriving at the border of atom B, on account of the tendency to
maintain its original momentum and that the instantaneous separation
distance $d_{AB}(t)$ between atom A and B is sufficiently small, the
electron can thus go into the orbital of atom B with certain probability,
moving anticlockwise around atom B\cite{argue1}. Put alternatively, the
electron has two choices, i.e., it either moves around atom A in the
original direction, or transfers to its nearest neighboring atom B and moves
around it in the opposite direction. Similarly, when the electron arrives at
the edge of atom C, it can go into the orbital of atom C and moves again
clockwise around it, and anticlockwise around atom D $\cdots $ The electron
thus transfers from atom A to D smoothly, without being reflected or
scattered and consequently losing its energy, which means non-resistivity
macroscopically.\vspace{5cm}

\begin{center}
\textbf{figure 2}
\end{center}

It should be emphasized that it is the small average separation distance $%
d_{AB}(t)$ between lattice atoms that allows conduction electrons to move in
such a path of semi-circles with arrows from one atom to another shown in
Fig.2. In other words, the electrons will naturally move in this interesting
path among lattice atoms and superconductive state forms, as long as the
vibrating lattice atoms are so close to each other that it matches(is
harmonic to) the energy and momentum of free electrons in the bulk material.
It should also be noted that a conductor will exhibit superconductivity as
long as there are in the bulk material such continuous passageways shown in
Fig.2, no matter how zigzag they are and how few there are, even only one
atom chain.

As we have arrived at the conclusion that the average separation distance
between atoms is crucial for a conductor to transit from a normal to a
superconductive state, now we have to consider a crucial problem: how to
describe this average distance? We introduce a functional $S$ to do this as
follows. 
\begin{equation}
S=\frac{\dsum\limits_{<i,j>}\frac 1{t^{\prime }}\int_0^{t^{\prime
}}d_{ij}^2(t)dt}{\dsum\limits_{<i,j>}}  \label{1}
\end{equation}
where $d_{ij}(t)$ is the instantaneous separation distance between nucleus $%
i $ and $j$ at time $t$, and the sum extends over all nearest-neighbor pairs
of nuclei. The integral in Eq. (\ref{1}) represents the time average of $%
d_{ij}(t)$. The function $d_{ij}(t)$ is related to the kinetic energy and
interaction potentials of the nuclei, while their kinetic energy is related
largely to the bulk temperature of the material\cite{argue2}. This means
that the functional $S$ is a function of the temperature of the material and
the potential between lattice atoms. Thus one can write 
\begin{eqnarray}
S &=&S(E_k(T),v)  \nonumber  \label{2} \\
&=&S(T,v)  \label{2}
\end{eqnarray}
where $E_k(T)$ is the average kinetic energy of lattice atoms in a bulk
material, $T$ is the temperature of the bulk material and $v$ is the average
potential between lattice atoms.

With this physical picture of superconductivity we can now proceed to
explain the experimental facts of superconductors. We exemplify only a few
basic ones, while others can also be interpreted without difficulty.

\textit{Non-resistivity. }As aforementioned, due to the sufficiently small
average separation distance between lattice atoms, conduction electrons can
transmit from atom to atom without being scattered and thus no macroscopic
resistance raised.

\textit{Meissner effect. }According to Ampere's molecular current theory of
magnetism, a bulk material's magnetic field is formed by its atomic
currents, i.e., all valence electrons moving around their nuclei in the same
direction. However, in this model of superconductivity, the number of
electrons moving clockwise equals that of anticlockwise. Consequently, the
macroscopic average magnetic field in a bulk superconductor is zero. It
should be reiterated that it is the small separation distance between nuclei
and appropriate momentum or energy of conduction electrons that makes the
conduction electrons move in such a semi-circle trajectory with arrows shown
in Fig.2. Any sufficient increase in energy and momentum of an electron or
in the average separation distance between lattice atoms will destroy this
subtle transferring pattern. Therefore, when the applied external magnetic
field is strong enough , the interaction between the electrons and the
magnetic field will break this harmony between conduction electrons and
lattice atoms, and consequently the material loses superconductivity and
becomes again normal state.

\textit{Frequency dependent electromagnetic behavior. }As the superelectron
absorbs enough energy from the electromagnetic field, the pattern of
electronic movement is also undermined. The analysis is similar to that of
Meissner effect.

An interesting and exciting conclusion can be drawn from this physical
picture of superconductivity that electrons move around nuclei in reverse
direction to each other one by one. Taking into account that valence
electrons in a ferromagnetic substance move in approximately the same
direction\cite{argue3}, we can deduce that given an other kind of atoms be
successfully sandwiched between ferromagnetic atoms one by one, i.e.,
replace all the anticlockwise atoms in Fig.2, with all their electrons
moving exactly in reverse direction to their neighboring ferromagnetic
atoms, a superconductor constructed this way, is expected most likely to
have a very high transition temperature.

In conclusion, we have proposed a novel mechanism of superconductivity
applicable to ``low temperature'' and ``high temperature'' superconductors.
Anderson suggested in 1987 that the new high $T_c$ materials may arise from
purely repulsive interactions\cite{Anderson}, which was motivated by the
fact that the superconductivity seems to originate from doping an otherwise
insulating state. However, this can be interpreted by the present model as
that the doped atoms act as bridges across the intrinsic atoms, shortening
the average separation distance $S$ defined by Eq.(\ref{1}), and thus
increases transition temperature of the doped material.


\begin{thebibliography}{99}
\bibitem{Onnes}  H. K. Onnes, Comm. Leiden 120b (1911), H. K. Onnes; \\Comm.
Leiden, Suppl. Nr 34 (1913).

\bibitem{BCS}  J. Bardeen, L. N. Cooper, and J. R. Schrieffer, Phys. Rev. 
\textbf{108}, 1175 (1957).

\bibitem{Bednorz}  J. G. Bednorz, and K. A. M\"{u}ller, Z. Physik B \textbf{%
64}, 189 (1986).

\bibitem{hi-tc}  A. Zettl, W. A. Vareka \& X. -D. Xiang, in \textit{Quantum
Theory of Real Materials, }edited by James. R. Chelikowshy \& Steven G.
Louie (Kluwer Academic Publishers, 1996), p. 425-448.

\bibitem{C.W.Chu}  Paul Ching-Wu Chu et al., in \textit{Quantum Theory of
Real Materials,} edited by James. R. Chelikowshy and Steven. G. Louie
(Kluwer Academic Publishers, 1996), p.411-489, and references therein.

\bibitem{Anderson}  P. W. Anderson, Science \textbf{235}, 1196 (1987).

\bibitem{atom}  see, for example, J. L. Mayo, \textit{Superconductivity: the
threshold of a new technology}, TAB Book Inc.(1988), p.18-20.

\bibitem{argue1}  This may be considered to conflict with standard electron
gas model which is characterized by Bloch wavefunction. However, it is noted
that Bloch's wavefunction is obtained in the approximations that the lattice
atoms are perfectly periodic and at rest, which is not the realistic
situation. The assumption $V(x)=V(x+na),$ upon which Bloch wavefunction is
obtained, will not hold in a real, even perfect crystal at finite
temperature, especially at high temperature. We argue that an electron in
any bulk material must move around some certain nucleus with certain
probabilities, maybe this one at this time, that one at that time, but it
``rotates'' around nuclei.

\bibitem{argue2}  One major factor affecting the potential between lattice
atoms is external pressure, which has been studied extensively. For detail,
see, for example, J. S. Schilling and S. Klotz, in \textit{Physical
Properties of High Temperature Superconductors III,} edited by Donal M.
Ginsberg (World Scientific, 1992), p.59-157.

\bibitem{argue3}  In a pure ferromagnetic metal, e.g., iron, its conduction
electrons have a strong tendency to rotate around their nuclei at the same
direction. With the present model of superconductivity, it is not difficult
to understand why ferromagnetic elements have no superconductive transition
temperatures.
\end{thebibliography}
\end{document}